\documentclass{iopart}

\usepackage{epsfig}
\usepackage{amscd}
\usepackage{graphicx,xspace}
\input{epsf}

\include{latexcommands}

\def\be{\begin{equation}}
\def\ee{\end{equation}}

\def\bea{\begin{eqnarray}}
\def\eea{\end{eqnarray}}

\def\s{\sigma}
\def\e{\epsilon}
\def\l{\lambda}

\def\k{\kappa}

\begin{document}

\title[Off-diagonal correlations in 1D anyonic models]{Off-diagonal 
correlations in one-dimensional anyonic models: A replica approach}

\author{Pasquale Calabrese${}^{1}$ and Raoul Santachiara${}^{2}$}

\address{$^{1}$Dipartimento di Fisica dell'Universit\`a di Pisa and INFN, 
Pisa, Italy}

\address{$^{2}$ Laboratoire de Physique Th\'eorique et Mod\`eles
  Statistiques. Universit\'e Paris-Sud, F-91405 Orsay Cedex, France.}

\begin{abstract}

We propose a generalization of the replica trick that allows to calculate
the large distance asymptotic of off-diagonal correlation functions in
anyonic models with a proper factorizable ground-state wave-function.
We apply this new method to the exact determination of all the harmonic 
terms of the correlations of a gas of impenetrable anyons and to the 
Calogero Sutherland model. 
Our findings are checked against available analytic and numerical results.

\end{abstract}

\maketitle

\section{Introduction}

Generalized anyonic statistics, which interpolate continously between bosons 
and fermions, are considered one of the most remarkable breakthrough of 
modern physics.
In fact, while in three dimensions particles can be only bosons or fermions, 
in lower dimensionality they can experience exchange properties intermediate 
between the two standard ones \cite{anyons}. 
In two spatial dimensions, it is well known that fractional braiding
statistics describe the elementary excitations in quantum Hall effect, 
motivating a large effort towards their complete understanding. 

In one dimension, anyonic statistics are described in terms of fields that 
at different points ($x_1 \neq x_2$) satisfy the commutation relations   
\bea
\Psi_A^\dag(x_1)\Psi_A^\dag(x_2)&=&e^{i\k \pi\e(x_1-x_2)} 
\Psi_A^\dag(x_2)\Psi_A^\dag(x_1)\label{ancomm}\,,\\
\Psi_A(x_1)\Psi_A^\dag(x_2)&=&e^{-i\k \pi\e(x_1-x_2)} 
\Psi_A^\dag(x_2)\Psi_A(x_1)\,,
\nonumber
\eea
where $\e(z)=-\e(-z)=1$ for $z>0$ and $\e(0)=0$. $\k$ is called statistical
parameter and equals $0$ for bosons and $1$ for fermions.
Other values of $\k$ give rise to general anyonic statistics ``interpolating''
between the two familiar ones.

After the second life of one-dimensional gases started following their 
experimental realization in cold atomic setups \cite{1dexp}, there have been
several proposal to create and manipulate anyons \cite{par}, even 
in one dimension.
A few 1D anyonic models have been introduced and investigated 
\cite{k-99,bg-06,oae-99,g-06,ssc-07,zw-92,bg-06b,cm-07,pka-07,an-07,lm-99,it-99,kl-05,fibo,g-07,zw-07,o-07,sc-08,pka-08,hzc-08,dc-08,bgk-08,bcm-08}. 
In this paper we consider the calculation of off-diagonal correlation 
(or one-particle density matrix) for a system of $N$ particle 
\be
g_1^\k(x)=\rho_{N}^\k(x)=
\langle\Psi^{\k\dag}_A(x) \Psi^\k_A(0)\rangle\,,
\ee
of models that have a factorizable many-body ground-state wave function as
\be
\Phi_\k(x_1,\ldots, x_N) = C_N  \prod_{k<l} f(e^{2\pi ix_k/L}-e^{2\pi i x_l/L})
B_\k(x_k-x_l)\,,
\label{gs}
\ee
with $f(x)$ a generic even function, $C_N$ a normalization constant,
and 
 \be B_\k(x)=
  \cases{ e^{-i\pi \k}& $x<0$, \cr
          1  & $x>0$. }
  \label{1dbraiding2}
\ee
that completely specify the statistic.
Among these models, the more relevant are the anyonic generalization of
Tonks-Girardeau gas and of the Calogero-Sutherland model.

In this paper we propose a generalization of the replica trick that is able to
select the correct anyonic branch in the replicated space. 
The use of this replica method started in the context of the diagonal
correlation of the Calogero-Sutherland model \cite{gk-01}, has been later
extended to a gas of impenetrable bosons \cite{g-04,gs-06}, and finally to
the off-diagonal correlation of Calogero-Sutherland, but only for bosonic and
fermionic statistics \cite{agls-06}. 
It has been already proposed \cite{g-04} that in the replica approach one
deals with two functions of the replica number $n$, one for fermions and one
for bosons, or better a single function with two branches, a fermionic and a
bosonic ones that coincide only for integer $n$.
Here we push forward this interpretation and we claim that the replicated 
correlation has an infinite number of branches corresponding to the different
anyonic statistics. 
The apparent lack of rigor of this approach (that is shared by most of replicas
calculations starting from the celebrated Parisi mean-field solution of the
spin-glass \cite{sg}) can be justified a posteriori by the amazingly simple
final results that agree with all the available analytics and with the 
direct numerical calculations.   
The analytic continuation we propose in this paper allows us to obtain 
exact asymptotic expansions for large distances that are valid for the
off-diagonal correlation for general anyonic parameter $\k$. 
We also exploit the property that the anyonic correlation function satisfies
a Painlev\'e V differential equation \cite{sc-08}, that allows a rigorous
analytic check of the results obtained trough replicas. 
To further check the replica predictions, we perform a careful numerical
comparison of the finite-size off-diagonal correlation obtained via 
the Toeplitz determinant representation \cite{ssc-07,sc-08}.

The manuscript is organized as follows. In Sec. \ref{secTG} we introduce the
anyonic Tonks-Girardeau gas and the strategy for the replica calculation.
In Sec. \ref{secint}, starting from the example of a simple integral we show
how to select the correct anyonic statistic from the replicas.
In Sec. \ref{secrepbos} we review the replica calculation for the bosonic
impenetrable gas following Ref. \cite{g-04}.
In Sec. \ref{secrepany} we apply the replicas to the anyonic Tonks-Girardeau 
gas, showing that all the harmonic terms in the large distance expansion are
captured by the saddle-point approximation in the replica space. 
In Sec. \ref{secbeysp} we calculate the corrections to the saddle-point
approximation, both by perturbation theory and via the solution of the
Painlev\'e V equation. In Sec. \ref{secdet} we compute numerically the
off-diagonal correlation function, showing perfect agreement with the analytic
results. In Sec. \ref{secCS} we use the replica trick to obtain the 
correlation function in the fully anyonic Calogero-Sutherland model. 
Finally in Sec. \ref{secconcl} we draw our conclusions.

\section{The anyonic Tonks-Girardeau gas}
\label{secTG}

In first-quantization language, the Lieb-Liniger Hamiltonian is
\begin{equation}
H=-\sum_{i}^{N}\frac{\partial^2}{\partial x_{i}^{2}}+
2c\sum_{1\leq i<j\leq N}\delta(x_i-x_j),
\label{ALL}
\end{equation}
where the $N$-anyons ground-state function  exhibits the generalized symmetry
under the exchange of particles. 
For $\k = 0$ the model reduces to the bosonic Lieb-Liniger \cite{LL}, 
while for $\k=1$ to free fermions. The Tonks-Girardeau gas is obtained for
$c\to\infty$ and corresponds to impenetrable anyons.
In this limit, the ground-state wave function has the form of 
Eq. (\ref{gs}) with  $f(y)=|y|$.
This anyonic model has been explicitly studied in Refs. 
\cite{k-99,g-06,bg-06,ssc-07,cm-07,pka-07,pka-08,sc-08,dc-08}.

The anyonic one-particle density matrix for $N+1$ particles can be written 
\cite {ssc-07}, as
$N$-dimensional integral (in the variables $x_j/L=t_j$ and $t=x/L$)
\bea
\fl
\rho^\k_{N+1}(t)=\frac{1}{(N+1)!}\int_0^1d t_1\cdots\int_0^1d t_N
\nonumber\\
 \prod_{s=1}^N B_\k(t_s-t) |\sin[\pi (t_s-t)]||\sin[\pi t_s]|
  \prod_{1\leq i<j \leq N} 4\sin[\pi(t_j-t_i)]^2.
  \label{onepart_int}
 \eea
Using the identity
\begin{equation}
\fl \prod_{1\leq j<k \leq N} 4\sin[\pi(t_j-t_k)]^2=\prod_{1\leq j<k \leq
N} |e^{i 2 \pi t_k}-e^{i 2 \pi t_j}|^2\equiv |\Delta_N(\{t_i\})|^2,
\end{equation}
we can identify the second product in the integral
(\ref{onepart_int}) with the square of the absolute value of a
Vandermonde determinant. Thus the reduced density matrix is an
average over the distribution $|\Delta_N|^2$:
\be
\rho^\k_{N+1}(t)= 
  \left\langle \prod_{s=1}^N B_\k(t_s-t) |z-z_s| |1-z_s|
  \right\rangle_{|\Delta_N|^2}\,,
\ee
where $z_s=e^{i 2 \pi t_s}$ and $z=e^{i 2 \pi t}$.

In the case of bosons ($\k=0$) the reduced density matrix can be
obtained \cite{g-04} through a {\it replica trick} considering for a
positive integer $n$
 \be
  Z_{2n}^{\k=0}(t)=\left\langle \prod_{s=1}^N  (z-z_s)^{2n}
  (1-z_s)^{2n}
  \right\rangle_{|\Delta_N|^2}\,,
  \label{Repbos}
 \ee
and then $\rho^\k_{N+1}(t)$ is obtained from $Z_{2n}(t)$ by a
suitable analytic continuation to $n\to 1/2$. It has been shown by
Kurchan \cite{k-91} that to obtain the average of the absolute value from 
the replicated correlation, we have to 
require that the analytic continuation for any complex $n$
with positive real part should diverge as maximum as a power law.

How to modify such replica calculation to anyons when we deal with
averages over a function of unitary modulus but with arbitrary
phase? Our trick is to start by considering an anyonic parameter
that is a rational number, i.e. $\k= q/p$ with $p$ and $q$ integer.
We can then consider
 \be
  Z_{2np} (t)= \left\langle \prod_{s=1}^N  (z-z_s)^{2np}
  (1-z_s)^{2np}\right\rangle_{|\Delta_N|^2},
 \ee
that, after a proper analytic continuation, for $2np\to 1$ could
give the desired reduced density matrix. The ambiguity in this case
is obvious, in fact $Z_{2np}$ includes all the correlations
of anyons with statistical parameter $\k= q/p$ for any $q=1
\dots 2p$. This means that different analytic continuations will
give different anyonic correlations including bosonic and fermionic
ones. We will propose a general criterion that would allow to choose
the proper analytic continuation for any value of the statistical
parameter. To elucidate the meaning of our trick we start by
applying it to a trivial integral that contains all the essence of
the modified replica trick.

\section{The replica trick for an integral}
\label{secint}

Let us consider the integral
 \be
 I= \int_{-\infty}^{+\infty} dt \,|t| e^{-t^2}= 1\,,
 \ee
that has the replica representation
 \be
 I_{2n}= \int_{-\infty}^{+\infty} dt \,t^{2n} e^{-t^2}= \frac{1+e^{2\pi i n}}{2} \Gamma(1/2 +n)\,.
 \ee
Notice that $I_{1}=0$, while considering $n$ integer and
analytically continuing we have $I_{2n}^c=\Gamma(1/2+n)$, that leads
to $I_{2n}^{c}=1$, the desired result. This analytic continuation
selected the ``bosonic'' branch of the integral, giving the integral
with the absolute value, because it is the only analytic
continuation that does not diverge exponentially in the complex
plane for $n\to i \infty$, i.e. that satisfy Kurchan criterion for the
absolute value \cite{k-91} . Oppositely taking directly $I_{2n}$ we would
have obtained the ``fermionic'' branch.

Let us then modify the integral anyonically:
 \be
 I^{\k}= \int_{-\infty}^{+\infty} dt B_\k(t) |t| e^{-t^2}= \frac{1+e^{-i\pi q/p}}2\,,
 \ee
with $B_\k(t)$ given by Eq. (\ref{1dbraiding2}) and $\k= q/p$. The
proposed analytic continuation is
 \be
 I_{2np}= \int_{-\infty}^{+\infty} dt\, t^{2np} e^{-t^2}= \frac{1+e^{2i \pi n p}}{2} \Gamma(1/2
 +np)\,,
 \ee
where we loose any knowledge of the value of $q$, that is the
previous mentioned ambiguity. Considering $n$ and $p$ integers and
analytically continuing one would always get the bosonic branch. On
the other hand, one can play the following game
 \be
  I_{2np}=\frac{1+e^{-2i\pi n q }e^{2i\pi n (p+q)}}{2}
  \Gamma(1/2+np)\,,
 \ee
with $q,p,n$ integers. Considering the analytic continuation
 \be
  I_{2np}^{q}=\frac{1+e^{-2i \pi n q }}{2}
  \Gamma(1/2+np)\,,
 \ee
we get
 \be
  I_{1}^{q}=\frac{1+e^{-i \pi q/p }}{2}\,,
 \ee
that selects the desired anyonic branch. The reason why this simple
game works is that $I_{2np}^{q}$ is the only analytic continuation
which has the correct asymptotic behavior for large imaginary $n$
(and positive real part) that selects the correct branch. The
message is that the integral $I_{2np}$ contains all statistics from
bosons to fermions and a given one can be obtained just by selecting
the correct behavior at large imaginary $n$.

With this observation in mind, we can then use the result $Z_{2n}(t)$ for the
bosonic Tonks-Girardeau gas and for the Calogero-Sutherland model. 
Properly continuing $Z_{2np}(t)$, we easily get the anyonic result. 
Thus in the next section we review the bosonic
result and in the following one we generalize it to anyons.

\section{Bosons review}
\label{secrepbos}

All the content of this section is taken from the pioneering work of Gangardt
\cite{g-04}. We report this material here only to make this paper
self-contained. 

The replicated average for bosons Eq. (\ref{Repbos}) is (we 
recall $z=e^{2\pi i t}$)
\begin{equation}
\fl  \label{eq:circ_Zn_def}
  Z_{2n} (t) = \frac{1}{M_N(2n)} \int_{-1/2}^{1/2} d^N t
  |\Delta_N (e^{2\pi i t})|^2 \prod_{s=1}^N
  |1+e^{2\pi i t_s}|^{2n}|z+e^{2\pi i t_s}|^{2n},
\end{equation}
where we have normalized $Z_{2n}(1) = 1$, introducing the
constant $M_N(2n)$ (see Ref. \cite{g-04} for its
precise value, it is not essential for what follows). 
The integral (\ref{eq:circ_Zn_def}) has a dual representation \cite{fw-04}
by an integral over $n$ variables
\begin{equation}
  \label{eq:circ_duality}
  Z_{2n} (t) = \frac{z^{-Nn}}{S_{2n}} \int_0^1 d^{2n} x\, \Delta_{2n}^2
  (x) \prod_{a=1}^{2n} (1-(1-z)x_a)^N,
\end{equation}
where the normalization constant is given by the Selberg integral
\begin{equation}
  \label{eq:circ_selberg}
  S_{2n} = \int_0^1 d^{2n} x \,\Delta_{2n}^2 (x) = \prod_{a=1}^{2n}
  \frac{\Gamma^2(a) \Gamma(1+a) } {\Gamma(2n+a)}\,,
\end{equation}
In Eq. (\ref{eq:circ_duality}) the number of
particles $N$ appears only as a parameter. This representation
allows us to obtain the asymptotic expression for $Z_{2n}$ suitable
for analytic continuation in $n$. In the large $N$ limit the
integrand in (\ref{eq:circ_duality}) oscillates rapidly and the main
contribution comes from the endpoints $x_\pm=1,0$ which are the only
stationary points of the phase. We change variables near each
endpoint
 \begin{eqnarray}
   \label{eq:circ_chang_var}
   x_a&=&0+\frac{\xi_a}{N(1-z)}
,\qquad\qquad\qquad a=1,\ldots,l, \nonumber\\
   x_b&=&1-\frac{\xi_b}{N(1-z^{-1})}
   ,\qquad\qquad b=l+1,\ldots,2n ,
 \end{eqnarray}
The integrand in (\ref{eq:circ_duality}) simplifies in the large $N$
limit:
\begin{equation}
  (1-(1-z)x_a)^N\simeq \left\{
    \begin{array}{ll}
      e^{-\xi_a}, & a=1,\ldots,l\\
      z^N e^{-\xi_a}, & a=l+1,\ldots, 2n
    \end{array}\right.
  \label{eq:circ_spm}
\end{equation}
and  the integration measure including the Vandermonde determinant
is factorized as
\begin{equation}
  \label{eq:circ_vandermonde_fact}
  d^{2n} x\,\Delta^2_{2n} (x)
  = \frac{d^{l}\xi_a\,\Delta^2_{l} (\xi_a) \;d^{2n-l}\xi_b\,\Delta^2_{2n-l} (\xi_b)}{ [N(1-z)]^{l^2} [N(1-z^{-1})]^{(2n-l)^2}}.
\end{equation}
The  remaining integrals are calculated using \cite{metha,df,fz-96}
\begin{equation}
  \label{eq:lag_selberg}
 I_l (\lambda) = \int_0^\infty d^l \xi_a \Delta^2_l (\xi_a)
  \prod_{a=1}^l e^{-\lambda \xi_a} = \lambda^{-l^2} \prod_{a=1}^{l}
  \Gamma(a)\Gamma(1+a).
\end{equation}
Multiplying the contribution of each saddle point by the number of
ways to distribute variables we obtain the asymptotic of the
integral (\ref{eq:circ_duality}) as a sum of $2n+1$ 
terms \footnote{we corrected an error/typo irrelevant  in the limit 
$n\to1/2$ in Ref. \cite{g-04}, i.e. a factor that is one when $n\to 1/2$.}
\begin{equation}
  \label{eq:circ_Zn_2}
  Z_{2n} (t) = \prod_{c=1}^{2n} \frac{\Gamma (2n+c)}{\Gamma(c)} \sum_{l=0}^{2n}
   (-1)^{2n(n-l)}
   \frac{\left[F^l_{2n}\right]^2 z^{(N+2n)(n-l)}}
      {{(2X)^{l^2+(2n-l)^2}}},
\end{equation}
where we have introduced $X=N(z^{1/2}-z^{-1/2})/2i= N\sin\pi t$ and
the factors $F^l_{2n}$
\bea
 F_{2n}^l &=
    \prod_{a=1}^l \frac{\Gamma(a)}{\Gamma{(2n-l+a)}}=
    \frac{G(l+1) G(2n-l+1)}{G(2n+1)}\,,
 \eea
where $G(x)$ is the Barnes $G$-function
\begin{equation}
\fl G(z+1)=(2 \pi)^{z/2} e^{-(z+(\gamma_E+1)z^2)/2}
\prod_{k=1}^{\infty}(1+z/k)^k e^{-z+z^2/(2k)},
\end{equation} 
$\gamma_E$ is the Euler constant, $G(1)=G(2)=1$, and $G(3/2) = 1.06922\dots$.
Using the Barnes function is particularly useful to get the analytic 
continuation easily.
Notice that also at this level, if we set $2n=1$ in
Eq. (\ref{eq:circ_Zn_2}) we still get the fermion result 
$g_1^{\k=1}(x)= \sin((N+1)\pi t)/ ((N+1) \sin\pi t)$.

For $n$ integer, we can extend the sum from $0\leq l\leq 2n$ to all the 
integers, because $F_{2n}^l$ vanishes for $l<0$ and $l>2n$ (it is a simple
properties of the Barnes $G$ function inherited from the $\Gamma$).
While this is an innocuous change for integer $n$, it is the fundamental step 
that will allow for replica symmetry breaking, because for general values 
of $n$ (i.e. non integers) $F_{2n}^l$ are non zero for any $l$ and the sum 
results in a true infinite series. 
Changing the summation variable to $k=l-n$  we obtain finally
\begin{equation}
\fl  \label{eq:circ_Zn_3}
  Z_{2n} (t) = \frac{\prod_{c=1}^{2n} \Gamma (2n+c)/\Gamma(c)}
  {(2N |\sin \pi t|)^{2n^2}} \left(G^4(3/2)+2\sum_{k=1}^\infty (-1)^{2nk}
    \left[F^{k+n}_{2n}\right]^2 \frac{\cos \left[2k\, \pi (N+2n) t\right] }
    {(4N^2\sin^2\pi t)^{k^2}}\right).
\end{equation}
This apparently useless shift is the crucial point where the replica
symmetry is finally broken because it selects the zero mode. 
Now we are in a position to take the limit $n\to 1/2$  which results
in the desired harmonic expansion of the one-body density matrix
\begin{equation*}
\fl \rho_{N+1}(t)
  =  \frac{1}{|2N \sin \pi t|^{1/2}}
    \left[\sum_{k=-\infty}^\infty
  (-1)^{k} [G(3/2+k) G(3/2-k)]^2
  \frac{\cos 2 k\,\pi (N+1) t}{|2 N\sin\pi t|^{2k^2}}\right].
\end{equation*}
The form of this expansion has been firstly predicted by Haldane \cite{h-81},
but the complete knowledge of all the amplitude has been only obtained by
means of replicas \cite{g-04}.

\section{The anyons}
\label{secrepany}

In the calculation of $Z_{2np}$ everything proceeds like in the
calculation for boson, leading to the replacement $n\to np$ in the
replicated correlation (\ref{eq:circ_Zn_2})
 \be
\fl  Z_{2np} (t) = \prod_{c=1}^{2np} \frac{\Gamma (2np+c)}{\Gamma(c)}
\sum_{l=0}^{2np}   (-1)^{2np(np-l)}
   \frac{\left[F^l_{2np}\right]^2 z^{(N+2np)(np-l)}}
      {{(2X)^{l^2+(2np-l)^2}}}.
 \ee
The correct analytic continuation to get the anyonic branch with
$\k=q/p$ is obtained by selecting the behavior at large imaginary
$n$. As for bosons, this is obtained by choosing the zero mode in
the sum over $l$ and then by extending the sum over all integers.
Thus we set
 \be
  np-l=nq+m\,,
 \ee
that, amazingly, is the only change we need to describe anyons from the
bosonic formulae. With this substitution we get
 \be
\fl  Z_{2np}^q (t) = \prod_{c=1}^{2np} \frac{\Gamma (2np+c)}{\Gamma(c)} \;
   \sum_{m=-\infty}^{\infty}
   (-1)^{2np(nq+m)}
   \frac{\left[F^{np-nq-m}_{2np}\right]^2 z^{(N+2np)(nq+m)}}
      {{(2X)^{(np-(nq+m))^2+(np+(nq+m))^2}}},
 \ee
that is ready for the analytic continuation $2np\to 1$, obtaining
 \be
\fl  \rho^{q/p}_{N+1} (z) =
   \sum_{m=-\infty}^{\infty}
   (-1)^{(q/2p+m)}
   \frac{\left[F^{1/2-q/2p-m}_{1}\right]^2 z^{(N+1)(q/2p+m)}}
      {{(2X)^{1/2 +2 (m+ q/2p)^2 }}},
 \ee
where only the ratio $\k= q/p$ appear and so it can be generalized
to any $\k$ not only ratio of integers, giving (restoring $z=e^{i2\pi x/L}$)
\be
 \fl  \rho^{\k}_{N+1} (x) =
   \sum_{m=-\infty}^{\infty} (-1)^{(\k/2+m)}
 \frac{\left[G(\frac32-\frac\k2-m) G(\frac32+\frac\k2+m)\right]^2 
   e^{i \pi x (\k+2m)(N+1)/L}}
      {{(2 N |\sin \pi x/L| )^{1/2 +2 (m+ \k/2)^2 }}}.
\ee

According to bosonization (or harmonic fluid approach), any
anyonic off-diagonal correlation in the thermodynamic limit 
$N\to\infty$ with $\rho_0=N/L$ kept constant, 
admits the expansion \cite{cm-07}
\be
g_1(x)=
\rho_0 \rho_\infty \sum_{m=-\infty}^{\infty} b_m
\frac{e^{i(2m+\k)\pi\rho_0 x}e^{-i (m+\k/2)\pi}}{
(\rho_0 d(x))^{\frac{(2m+\k)^2}2 K+\frac1{2K}}}\,.
\label{main}
\ee
Here $d(x)=|L\sin (\pi x/L)|$, $b_m$ are unknown non-universal amplitudes, 
while $K$ is the universal Luttinger liquid exponent equal to $1$
for Tonks-Girardeau (we fixed $b_0=1$).
The two expansions agree perfectly,
confirming the correctness of the result. Furthermore, we obtained  an exact
expression for all the coefficients in the harmonic expansion 
\be \fl
\rho_\infty 
b_m=  \frac{\left[G(\frac32-\frac\k2-m) G(\frac32+\frac\k2+m)\right]^2}
      {{2^{1/2 +2 (m+ \k/2)^2 }}}\quad {\rm with}\;
\rho_\infty=\frac{\left[G(\frac32-\frac\k2) G(\frac32+\frac\k2)\right]^2}{
\sqrt2}\,.
\ee

The leading term $\rho_\infty$ has been already obtained from
Fisher-Hartwig conjecture applied to the Toeplitz determinant
\cite{ssc-07,sc-08}, and the two results agree. All the other terms are
new. It is important to write down the ratio of two consecutive amplitudes 
\be
  \frac{b_{m+1}}{b_m}= 
-\left[\frac{\Gamma(3/2+\k/2+m)}{\Gamma(1/2-\k/2-m)}\right]^2 4^{1+2\k+2m}\,,
 \ee
The first corrections to the leading behavior are given by the terms
with $m=\pm1$, that satisfy
 \be
  b_1 b_{-1} = \frac1{256} (1-\k^2)^2\,.
\label{b1bm1}
 \ee
We will see that the same result also follows from the Painlev\'e V equation, 
given a further check of the replica approach. 

\section{Beyond the saddle points}
\label{secbeysp}

The formula just derived gives the leading term in $1/N$ of the
desired correlation function. As well known, this does not give {\it all}
the dominating terms in the thermodynamic limit, 
defined as $N\to \infty$ with $N \pi x/L= k_F x$ taken
constant. 
One then needs to calculate corrections to the previous terms. 
A systematic way to get these corrections is to perform the perturbation
theory close to the saddle points order-by-order considering in Eq. 
(\ref{eq:circ_chang_var}) the full expansion in $\xi_{a,b}$ close to
$x_{\pm}=0,1$, as done in Ref. \cite{g-04} for bosons.
This approach is very general and straightforward and gives all the $1/N$
corrections, but unfortunately becomes soon 
computationally cumbersome increasing the order. 
For this reason, we only report the first-order computation in the next
subsection. 
Alternatively, we can take a different approach based on the
solution of the Painlev\'e V equation, that gives very easily all the 
terms in the expansion, but only in the thermodynamic limit.

\subsection{First order in perturbation theory}

In Ref. \cite{g-04} the first order correction in $1/N$ to $Z_{2n}(t)$ has
been explictly calculated (see Eq. (65) there).
Using this result and simply posing $2n\to 2np$ and then performing
the shift $np-l=nq+m$, we get for the first order term (we recall that
$X=N\sin \pi t$ so it is $O(N)$)  
\bea
\fl\frac{i}X (np-nq-m)(np+nq+m) [(np+nq+m) e^{i\pi x}-(np-nq-m)e^{-i\pi x}]
\nonumber\\
-\frac1N [(np+nq+m)^3+(np-nq-m)^3]\,,
\eea
taking $2np\to 1$ and then considering only the zero mode $m=0$ (higher modes
give rise to higher corrections in $1/N$), we get 
\be
-\frac1{2N}(1+\k^2)+i\frac{1-\k^2}4 \k \frac1{N\tan \pi x/L}\,.
\label{1ord}
\ee
Notice that for bosons and fermions ($\k=0,1$) the tan correction is absent: 
the role of this term in the replica approach is only understood while
studying anyons. 
The first term $-(1+\k^2)/2N$ just corresponds to the natural 
replacement $N \to N+1$ in the denominator of the leading term.

\subsection{Large distance asymptotic expansion via Painlev\'e V equation}

In Ref. \cite{sc-08} we have shown that $\rho_\k(t)$ for finite $N$ satisfies
second order differential equation of Painlev\'e VI type. 
The differential equation does not depend on the anyonic parameter, i.e. 
it is the same for any $\k$. What characterizes the statistic are the boundary
conditions at $t=0$ that do depend on the anyonic parameter.
It is well known that in the thermodynamic limit, the Painlev\'e VI equation
reduces to a Painlev\'e V (see e.g. \cite{metha,fw-02,ffgw-03})) that has been
firstly derived for bosons by Jimbo et al. \cite{jmms-80}
\bea
(x \s'')^2 + 4 (x \s' - 1 - \s) (x \s' + {\s'}^2 - \s)=0\,,\\
{\rm with} \qquad \s(x)=x \frac{d \log \rho(x)}{dx}\,.
\eea
To get a systematic expansion for large $x$ from this equation 
we need to impose the anyonic boundary conditions at $x=\infty$. 
We need the knowledge of the first two leading terms in the harmonic-fluid 
expansion $b_0$ and $b_1$. Actually $b_0$ gives only the overall  
normalization, in fact $\s(x)$ does not depend on it, while 
$b_{-1}$ is essential to fix all the signs (in fact an old sign error in
$b_{-1}=b_1$ for bosons \cite{vt-79} caused all the expansion in
\cite{jmms-80} to have the wrong signs. This has been corrected only in 
\cite{g-04}, see also \cite{ctw-80}).

\begin{figure}[t]
\includegraphics[width=\textwidth]{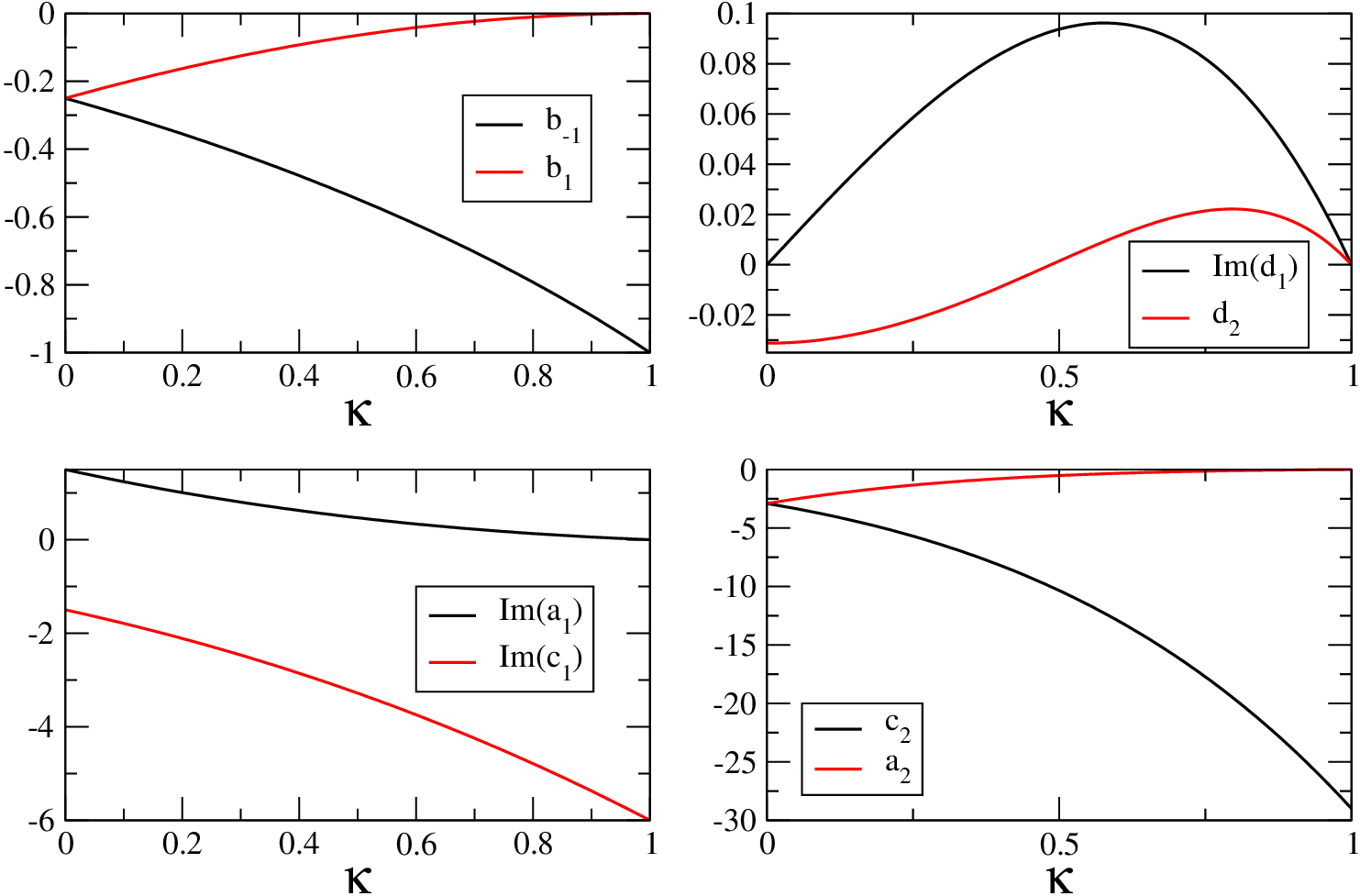}
\caption{Dependence on the anyonic statistic parameter $\k$ of the coefficients
of the large distance expansion of the off-diagonal correlators 
Eqs. (\ref{exp1}) and (\ref{expfs}).}
\label{figcoeff}
\end{figure}

In this subsection we set $k_F=1$ for simplicity in writing, 
the correct formulas are obtained by replacing $x\to k_F x$.
The complete asymptotic expansion of $\rho(x)$ is obtained by 
multiplying any term in the harmonic expansion (\ref{main}) by an analytic
function at $x=\infty$, i.e. 
 \bea
\fl\rho(x)=  \frac{\rho_\infty e^{i\k x}}{x^{1/2+\k^2/2}}
   \left[\left(1+\frac{d_1}x +\frac{d_2}{x^2}+\dots\right)
    +b_{-1} \frac{e^{-2 i x}}{x^{2-2\k}}\left(1+\frac{a_1}x
 +\frac{a_2}{x^2}+\dots\right)
\right.\nonumber\\ \left.
    +b_{1} \frac{e^{2 i x}}{x^{2+2\k}}\left(1+\frac{c_1}x +\frac{c_2}{x^2}+\dots\right)
    +\dots \right]\,.
\label{exp1}
 \eea
Form this we have the large distance expansion of $\sigma$ that fixes the
boundary conditions:
\be
\lim_{x\to\infty} \sigma(x)=i\k x -\frac{1+\k^2}2 -2 i b_{-1} e^{-2i x}
x^{-1+2\k}+ O(x^o)\,,
\ee
where $o$ is some not better specified exponent that will follow from the
equation. 
Plugging this expansion in the Painlev\'e equation 
one can get iteratively all the various coefficient $a_i,b_i,c_i,d_i$. 
The first term in this expansion is Eq. (\ref{b1bm1}), that provides a further
consistency check of the replica approach. 
After long and tedious algebra we got the first coefficients:
 \bea
  d_1&=& i\k\frac{1-\k^2}4\,,\\
  c_1&=& -i \frac{(1+\k)(2+\k)(3+\k)}4\,,\\
  a_1&=& i \frac{(1-\k)(2-\k)(3-\k)}4\,,\\
  d_2&=& \frac{1-\k^2}{32}(\k^4+4\k^2-1)\,,\\
  a_2&=& -\frac1{32} (3-\k)(1-\k)(31-48\k+28\k^2-8\k^3+\k^4)\,,\\
  c_2&=& -\frac1{32} (3+\k)(1+\k)(31+48\k+28\k^2+8\k^3+\k^4)\,,
 \eea
that for $\k\to0$ reproduce the known result for bosons \cite{g-04,jmms-80}. 
Notice that $d_1$ reproduces the result from first order
perturbation theory. 
In Fig. \ref{figcoeff} we report all these coefficients as function of $\k$. 
Notice that since at the fermionic point all the terms but two must vanish we
have $a_m=d_m=0$ for any $m$, while the vanishing of the other terms is
ensured by $b_1=0$. At the bosonic point instead the reality of the correlator
imply $a_i=c_i^*$ that is satisfied by our expressions.

\section{The comparison with numerics}
\label{secdet}

In this section we show how our expansions perfectly fit with the numerical
data obtained by the exact evaluation of the off-diagonal correlators at 
finite $N=L$ through the evaluation of the Toeplitz determinants 
derived in Refs. \cite{ssc-07,sc-08}.

\begin{figure}[t]
\includegraphics[width=\textwidth]{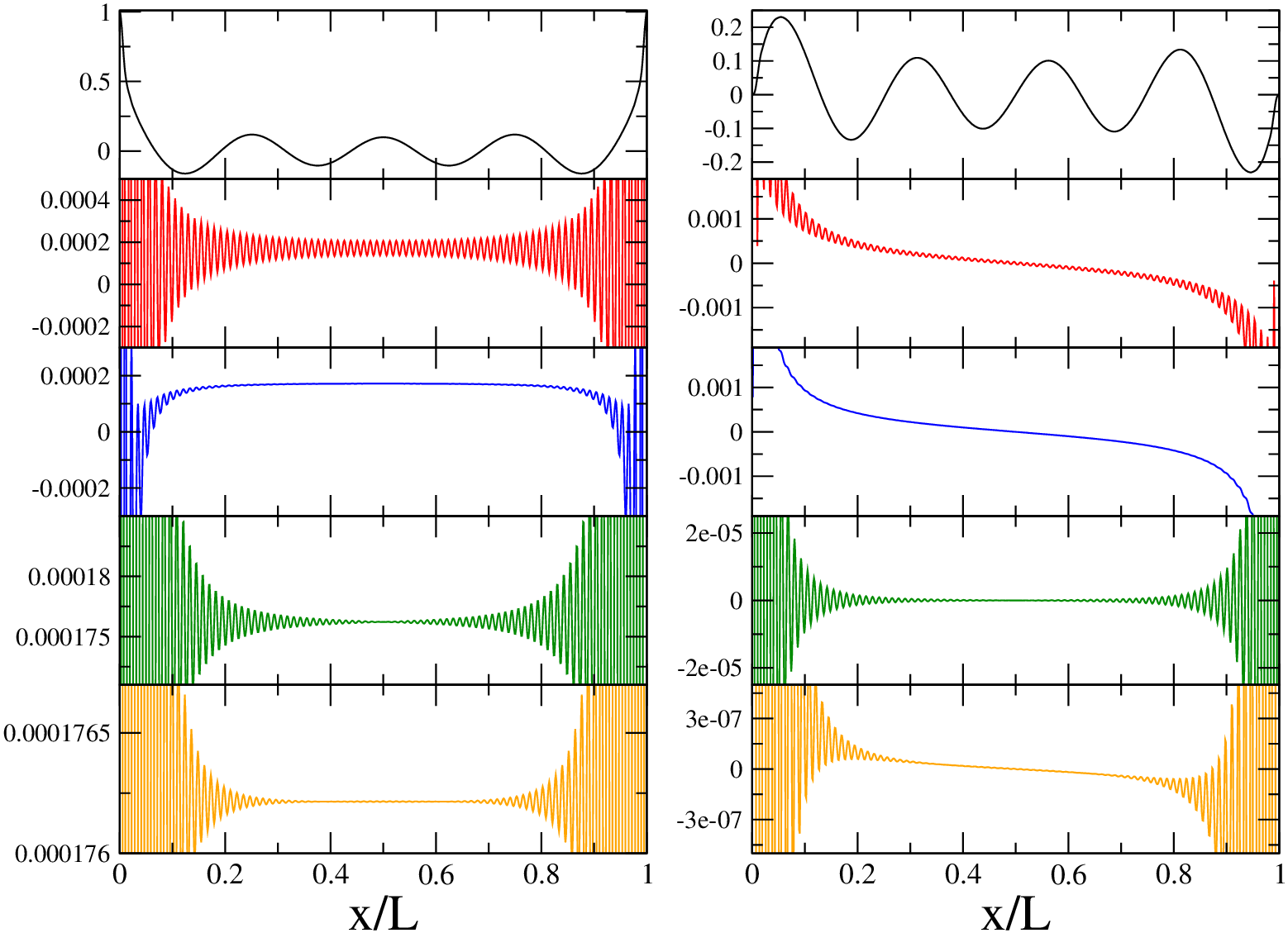}
\caption{Off-diagonal correlation function for $\k=0.1$ and $N=81$. 
Top panels: Real (left) and imaginary (right) parts of $\rho^{0.1}_{81}(t)$.
Second rows: $\rho^{0.1}_{81}(t)/\rho_{FH}(t)-1$. 
Third rows: Subtraction of the first two harmonic terms $b_{\pm1}$.
Fourth rows: Further subtraction of the analytic contributions $d_{1,2}$.
Bottomost panels: Further subtraction of the analytic corrections to the 
harmonic terms $a_1$ and $c_1$.}
\label{fig0.1}
\end{figure}

At finite volume and unit density, the expansion can be written as 
\bea
\fl\rho(x)&=&  
\frac{\rho_\infty e^{i\k x}}{[N \sin \pi x/L]^{\frac12+\frac{\k^2}2}}
   \left[1[1+O(N^{-2})]+\frac{d_1}{N \tan \pi x/L} +\frac{d_2}{[N \sin \pi x/L]^2}
+O(N^{-3})\right.\nonumber\\ \fl &&\left.
    +b_{-1} \frac{e^{-2 i N x}}{[N\sin \pi x/L]^{2-2\k}}
  \left(1+\frac{a_1}{N \tan\pi x/L} +\frac{a_2}{[N\sin\pi x/L]^2}
+O(N^{-3})\right)
\right.\nonumber\\ \fl &&\left.
    +b_{1} \frac{e^{2 i Nx}}{[N\sin \pi x/L]^{2+2\k}}
  \left(1+\frac{c_1}{N\tan\pi x/L} +\frac{c_2}{[N\sin\pi x/L]^2}+O(N^{-3})
 \right)
    \right.\nonumber\\ \fl && \left. + {\rm higher} \; {\rm harmonics} 
  \frac{}{}\right]\,.
\label{expfs}
\eea 
Notice that we have used in the leading term $N$ instead of $N-1$ 
to cancel the $1/N$ correction to the constant in the first line, according to
Eq. (\ref{1ord}).
It has been shown already in Ref. \cite{ssc-07} that only the leading term
in this expansion (obtained from Fisher-Hartwig conjecture) almost perfectly 
describe $\rho(x)$ for all $\k<1/2$. In Ref. \cite{cm-07} it has been argued
that for larger value of $\k$ the first harmonic term (i.e. $b_{-1}$) is
fundamental to have a good description of the asymptotic results because
its power  $2-2\k$ gets closer to zero. Including this term, we have a 
perfect description of the asymptotic behavior \cite{cm-07}, but the exact 
value of the amplitude $b_1$ has been obtained only here.

\begin{figure}[t]
\includegraphics[width=\textwidth]{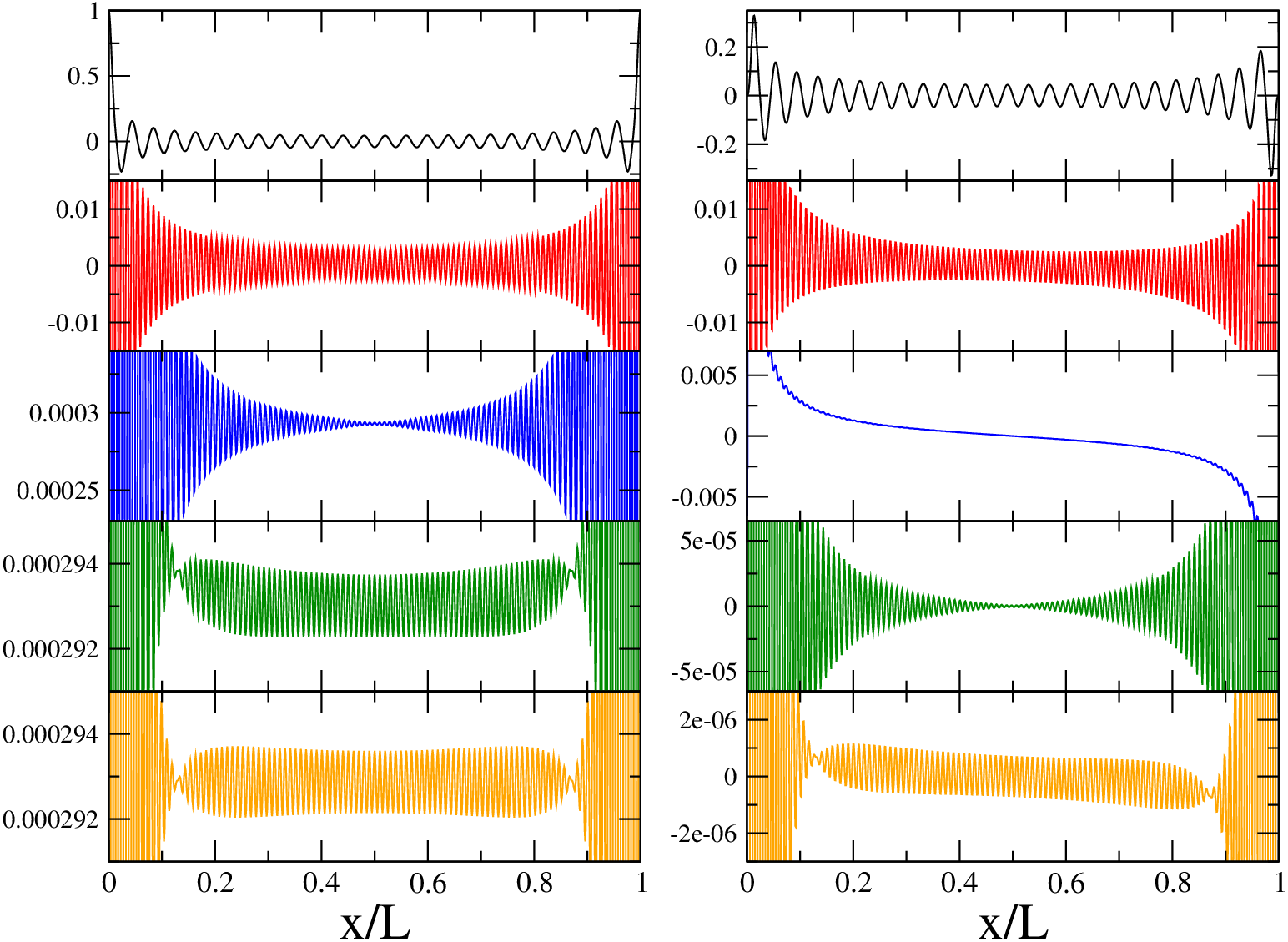}
\caption{Off-diagonal correlation function for $\k=0.5$ and $N=101$. 
Top panels: Real (left) and imaginary (right) parts of $\rho^{0.5}_{101}(t)$.
Second rows: $\rho^{0.5}_{101}(t)/\rho_{FH}(t)-1$. 
Third rows: Subtraction of the first two harmonic terms $b_{\pm1}$.
Fourth rows: Further subtraction of the analytic contributions $d_{1}$ (right)
and $a_1,c_1$ (left).
Bottomost panels: Further subtraction of the analytic corrections $d_2$ (left)
$a_1$ and $c_1$ (right).}
\label{fig0.5}
\end{figure}

\begin{figure}[t]
\includegraphics[width=\textwidth]{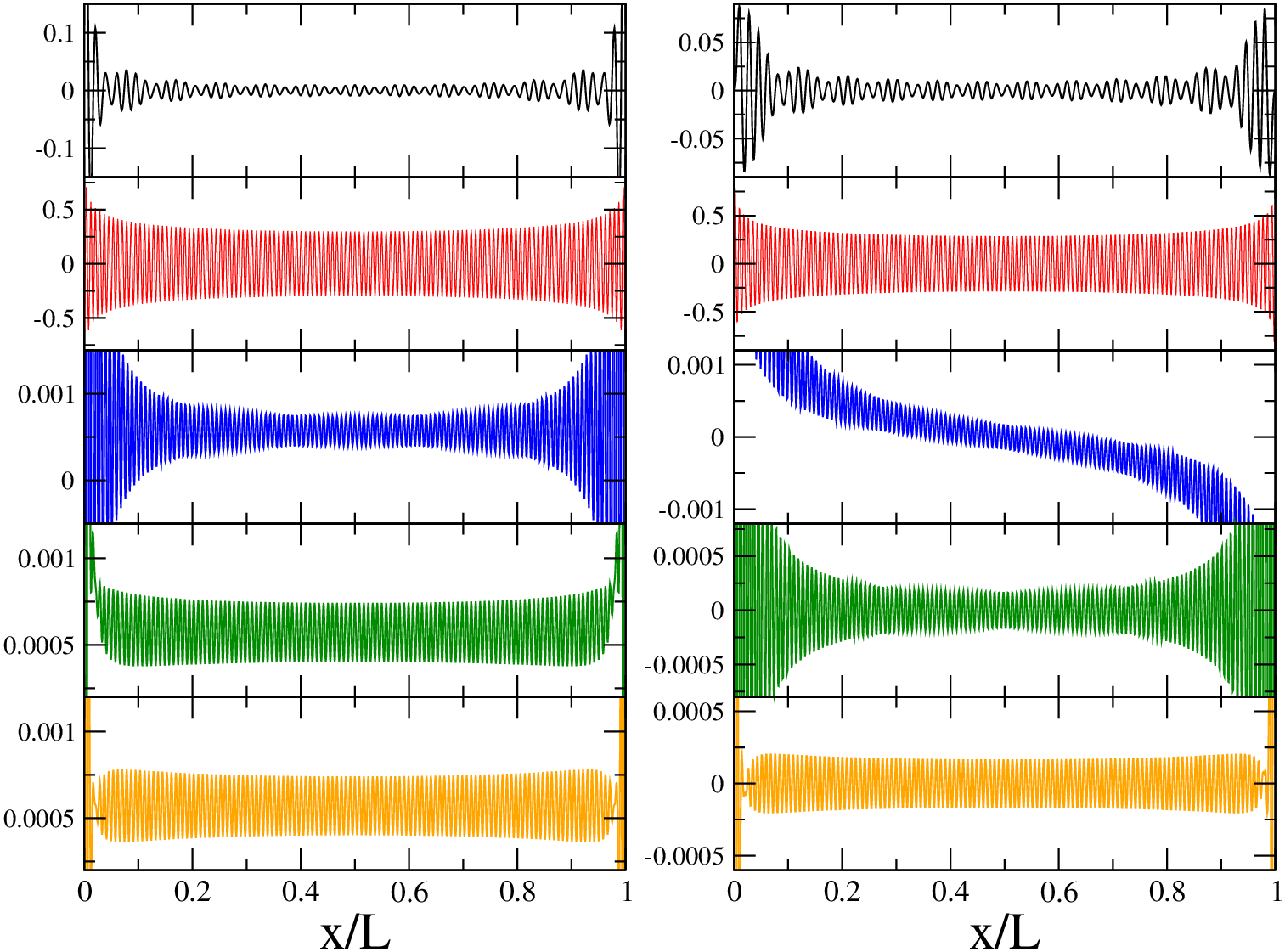}
\caption{As Fig. \ref{fig0.5} with $\k=0.9$ and $N=121$.} 
\label{fig0.9}
\end{figure}

Why wonder about further corrections? The main reason to care about them is
that they give rise to non-analytic terms in the momentum distribution
function at $k=\k+n k_F$ whose exact structure is characterized by the
coefficients $b_m$. Furthermore they give visible corrections to the
off-diagonal correlators if we carefully search for them. 
This is elucidated in the there figures \ref{fig0.1}, \ref{fig0.5}, and
\ref{fig0.9} where we report $\rho(x)$ for $\k=0.1$ and $N=81$, 
for $\k=0.5$ and $N=101$, $k=0.9$ and $N=121$ respectively.
In the three figures, the top panels show the real (left) and 
imaginary (right) parts of $\rho(x)$ directly calculated from the Toeplitz
determinant. 
In the second panels from the top we report real and imaginary parts of 
\be
\rho_{\rm corr}(x)=\frac{\rho(x)}{\rho_{FH}(x)}-1=
\frac{\rho(x)[N\sin\pi x/L]^{\frac12+\frac{\k^2}2}}{\rho_\infty e^{i\k x}}-1\,,
\ee
that should tend to zero for $N\to\infty$ for any $\k\neq1$ (at $\k=1$, the 
first harmonic $b_{-1}$ is of order $O(N^0)$).
We can in fact see that these are much smaller than the leading contribution
for $\k$ far from $1$, but become very relevant at $\k=0.9$ also at $N=121$.
Furthermore the second panels show very interesting oscillating behavior 
due to the first harmonic term. 
Thus, in order to kill the first oscillating behavior we subtract from 
$\rho_c(x)$ the first oscillating terms in Eq. (\ref{expfs}), i.e. 
the terms in $b_{\pm1}$. This subtraction is shown in the set third (from top)
of panels. The effect of this subtraction is rather peculiar. In fact 
while for $\k=0.1$ it mainly kills the oscillations, leaving the absolute
value almost unchanged because of the smallness of the exponent in 
$b_{\pm1}$, for larger $\k$ the absolute value goes down drastically, 
but strong oscillations are still present. The left oscillations are the 
second harmonics $b_{-2}$. 
In the imaginary part of what is left, one can clearly recognize a 
$1/\tan\pi x/L$ shape due to the term $d_1$ for any value of $\k$. 
In fact, by subtracting the $d_1$ term, as shown in the right fourth 
(from top) panels the value goes still drastically down of one- or two-order of
magnitude. 
Oppositely the real part at the third level has a shape depending on $\k$. 
This is very easily understood: For $\k<1/2$ the larger term left in the 
expansion is $d_2$, going always like $N^{-2}$, while for $\k>1/2$ 
it is the term in $a_1$ going like $3-2\k$. At $\k=1/2$ the two terms are 
both $N^{-2}$, but being $d_2$ very close to zero (see Fig. \ref{figcoeff}),
the most important term is $a_1$. 
For these reasons, in the left fourth panel is shown for $\k=0.1$ 
the subtraction of $d_2$ and in the other two cases the subtractions of $a_1$
and $c_1$. Evidently these subtractions are the correct ones and in fact 
decreases the considered value of at least one order of magnitude. 
Finally in the bottomost panel we report the final subtractions left of 
$a_1$, $c_1$ or $d_2$ depending on the case. A further reduction of the value
is observed. 
Clearly this procedure of subtraction can be repeated ad libidum, obtaining 
better and better agreement, but we preferred to stop at this level. 

A last comment concerns the absolute value of the real part of these
corrections, that is small, but non-vanishing, as evident from the plots. 
The $1/N^2$ contribution to this term can be easily obtained from second order
perturbation theory, as done for bosons in Ref. \cite{g-04}, but the
asymptotic nature of this expansion makes this calculation not very useful.

\section{The general anyonic Calogero Sutherland model}
\label{secCS}

The Hamiltonian of the Calogero-Sutherland model is given by a long-range 
pair interactions controlled by the coupling parameter $\lambda$:
\be 
\label{ham}
H=-\sum_{i=1}^N\frac{\partial^2}{\partial x_i^2}
+\lambda(\lambda-1)\sum_{i\neq j} \frac{\pi^2/L^2}{\sin^2(\pi(x_i-x_j)/L)}.
\ee
The ground state wave function of the Hamiltonian (\ref{ham}) was found in
\cite{s-71} and can be written in the form (\ref{gs}) with $f(y)=|y|^\l$
with normalization constant 
\begin{eqnarray}
C^2_N (\lambda) = \frac{1}{L^N}\frac{\Gamma(1+\lambda)^N}{\Gamma(1+\lambda N)}.
\end{eqnarray}
This model has always been considered only in the case of bosonic and 
fermionic statistics, or in the anyonic ones arising from having an
analytic ground-state function, that implies $\k=\l$. 
While thermodynamic properties and diagonal correlations do not depend on
$\k$, but only on $\l$, off-diagonal ones are a signature of the anyonic
statistics and we can consider general $\k\neq \l$ as pointed out by Girardeau
\cite{g-06}. 
The off-diagonal correlators are easily 
obtained with the same replica trick as before. 

The one-body density matrix ($t=2\pi x/L$) for $N+1$ particle in a ring is
\be
\fl  \label{eq:G1}
  g_1(x) = \frac{\Gamma(\lambda)\Gamma(1+\lambda N) }{2\pi
  \Gamma(\lambda(1+N))} \left\langle \prod_{j=1}^N 
  B_\k(t-\theta_j)|1-e^{i\theta_j}|^\lambda
  |e^{i t}-e^{i\theta_j}|^\lambda\right\rangle_{N,\lambda},
\ee
where the average is defined as
$$
  \left\langle f\left(e^{i\theta_1},
      \ldots,e^{i\theta_N}\right)\right\rangle_{N,\lambda}= \frac{\Gamma^N
      (1+\lambda)}{\Gamma(1+\lambda N)}\int_0^{2\pi}
      \frac{d^N\theta}{(2\pi)^N} \left|\Delta_N
      (\{e^{i\theta_i}\})\right|^{2\lambda} f\left(e^{i\theta_1},
      \ldots,e^{i\theta_N}\right),
$$
and $\Delta_N (\{z_i\})$ is the usual Vandermonde determinant.

To calculate the average in (\ref{eq:G1}) we use the replica trick, along the 
lines of the calculation in \cite{gk-01} and \cite{g-04}, modified to include 
the anyonic statistics. Namely, consider the function
\begin{eqnarray}
  \label{eq:Z_def}
  Z^{(\lambda)}_{2np}(t) = \left\langle \prod_{j=1}^N (1-e^{i\theta_j})^{2np}
  (e^{i t}-e^{i\theta_j})^{2np} \right\rangle_{N,\lambda},
\end{eqnarray}
but now the replica limit is $2np\to \lambda$.
The duality transformation \cite{fw-04}, which enables 
to re-express the $N$-dimensional integral (\ref{eq:Z_def}) depending on the
parameter $2np$ as a $2np$-dimensional integral depending on $N$ as a 
parameter is the same as for bosons \cite{agls-06}
\be
\fl  \label{eq:Z_dual}
  Z^{(\l)}_{2np} (t) = \frac{e^{-iNnp t}}{S_{2np}(1/\l)} \int_0^1 d^{2np} x
  |\Delta_{2np}(\{x_a\})|^{\frac{2}{\l}} \prod_{a=1}^{2np}[x_a
  (1-x_a)]^{\frac{1}{\l}-1} [1-(1-e^{i t})x_a]^N .
\ee
The duality $\lambda\leftrightarrow 1/\lambda$ becomes evident by comparing 
the power of Vandermonde determinants in the previous equations.
We put $Z^{(\l)}_{2np} (0)=1$, so the normalization
constant is given by the Selberg integral
\begin{equation*}
\fl  
  S_{2np}(1/\l) = \int_0^1d^{2np} x \; |\Delta_{2np}(x)|^{\frac{2}{\l}}
\prod_{a=1}^{2np} x_a^{\frac{1}{\l}-1} (1-x_a)^{\frac{1}{\l}-1} =
\prod_{a=1}^{2np} \frac{\Gamma^2\left(\frac{a}{\lambda}\right)
\Gamma\left(1+\frac{a}{\lambda}\right)}{\Gamma\left(1+\frac{1}{\lambda}\right)
\Gamma\left(\frac{2np+a}{\lambda}\right)}.
\end{equation*}
One can perform the sum over the saddle points at $x_\pm=0,1$
arriving to the same form as before 
\be
\fl  Z_{2np} (t) = \sum_{l=0}^{2np}
   (-1)^{2np/\l(np-l)}
   \frac{H^l_{2np}(\l) e^{it(N+2np/\l)(np-l)}}
      {{(2X)^{(l^2+(2np-l)^2)/\l}}},
\ee
but with a $\l$ dependent $H^l_{2np}(\l)$ factor given for integer $2np=m$ 
in Ref. \cite{agls-06} as
\be
H^l_m(\l)= \frac{\Gamma(m+1)}{\Gamma(l+1)\Gamma(m-l+1)}
\frac{I_l(\l) I_{m-l}(\l)}{S_m(1/\l)}\,,
\ee
with $S_m(1/\l)$ given above and $
I_l(\l)=(\prod_{a=1}^l\Gamma[a/\l]\Gamma[1+a/\l])/\Gamma[1+1/\l]^l$.

After trivial algebra we get
\be
H^l_m(\l)= 
\prod_{c=1}^m \frac{\Gamma(\frac{m+c}{\l})}{\Gamma(\frac{c}\l)}
\left[
\prod_{a=1}^l \frac{\Gamma(\frac{a}\l)}{\Gamma(\frac{m-l+a}{\l})}
\right]^2\,.
\ee
For $\l=1$ this formula reduces to Eq. (\ref{eq:circ_Zn_2}) whose analytic
continuation to complex $n$ can be written in terms
of Barnes G-functions. Unfortunately this is not possible for general $\l$
(special formulas can be found for integer $\l$ or $1/\l$, but they 
are not very illuminating). Not having this nice representation in terms of
special functions, the analytic continuation becomes less simple, 
but still straightforward and it is reported in the appendix.

\begin{figure}[t]
\includegraphics[width=\textwidth]{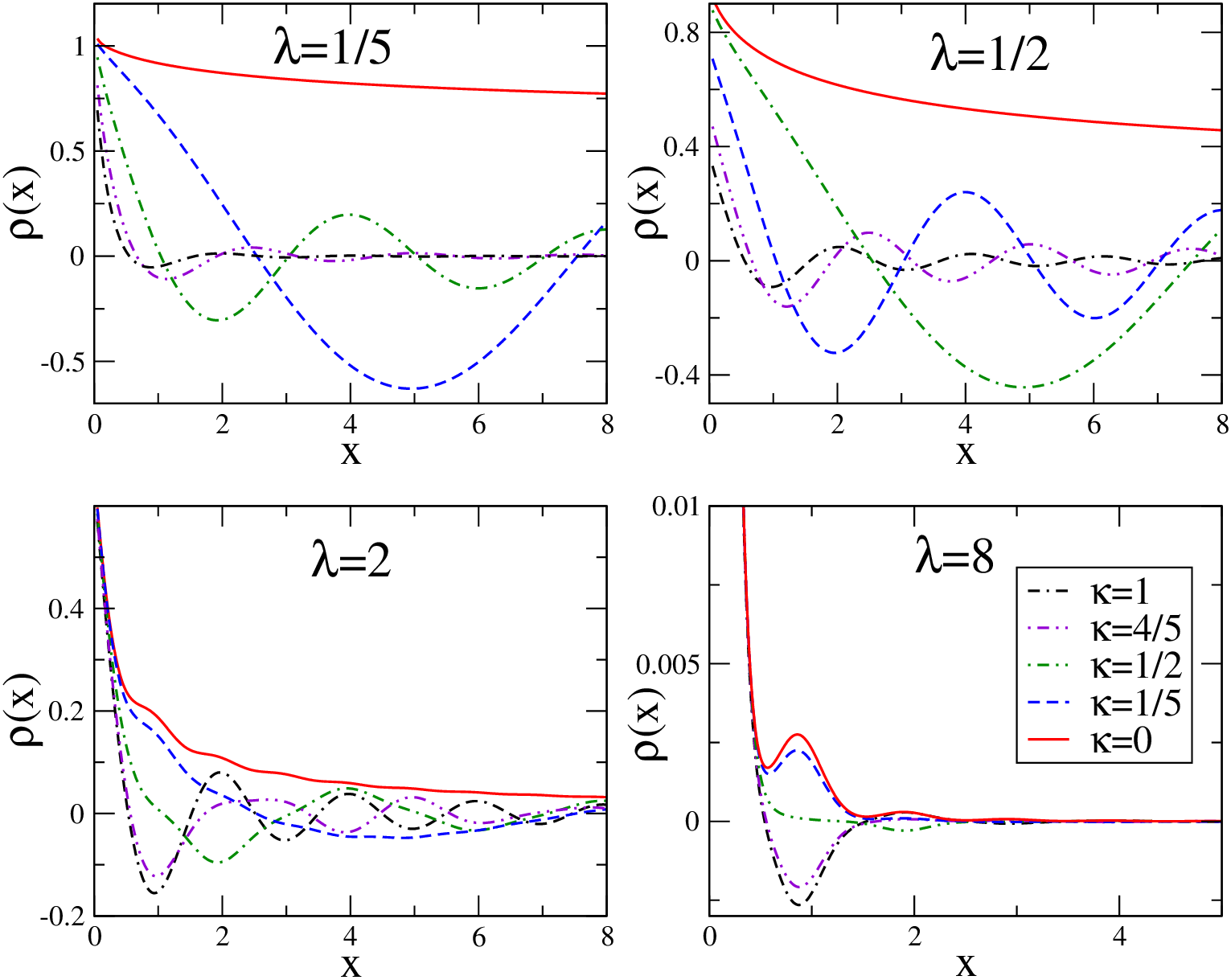}
\caption{One-body density matrix $g_1(x)$ for $N=51$, for values of $\l$ and 
several values of $\k$ (that are the same in the four plots). 
We consider the sum of the first few harmonics: $5$ harmonics are considered
for $\l>1$ and only one for $\l<1$, because as-well known, the 
asymptotic feature of the series are more severe for these value of $\l$.
Distances $x$ are measured in unit of the mean density.} 
\label{figCS}
\end{figure}

The anyonic statistics is obtained by setting $np-l=nq-m$
and so
\be
 \fl  g_1^{\k} (x) =
   \sum_{m=-\infty}^{\infty} (-1)^{(\k/2+m)}
 \frac{H_{\l}^{\l/2-\k/2-m}(\l)
   e^{i \pi x (\k+2m)(N+1)/L}}
      {{(2 N |\sin \pi x/L| )^{(\l/2 +2 (m+ \k/2)^2/\l )}}}\,,
\ee
where $H_1^{1/2-\k/2-m}(\l)$ is intended to be continued as explained in the
appendix.  

Explictly we get
\be \fl
\rho_\infty= \frac{(-1)^{\k/2}}{2^{\l/2+\k^2/(2\l)}}H^{\frac{\l-\k}2}_\l\,,
\qquad
\frac{b_{m+1}}{b_m}= - \left[
\frac{\Gamma(\frac12+ \frac\k{2\l}+\frac{m+1}{2\l})}{
\Gamma(\frac12-\frac\k{2\l}-\frac{m}{2\l})}\right]^2 
2^{2(2m+2\k+ 1)/\l}\,, 
\ee
and $b_0=1$.
This expression agrees with the general harmonic expansion (\ref{main}) 
with $K=\l^{-1}$ as well
known (the parameter $K$ do not depend on the statistic $\k$ and so it is the
same derived by Ha \cite{h-94} for $\k=\l$).
In Fig. \ref{figCS} we report the correlations obtained by summing few of the  
first terms in the harmonic expansion for several
values of the anyonic parameter $\k$ at fixed $\l$ to show the crossover from 
the fermionic to the bosonic curves. 
The extreme results $\k=0,1$ were already determined in Ref. \cite{agls-06}
(small manipulations are needed to show that they are identical),
where a perfect agreement with quantum Monte Carlo simulations has been shown.

This harmonic expansion has been previously calculated in specific 
cases (bosons and
fermions for generic $\l$ in \cite{agls-06} and for $\k=\l$ in \cite{h-94}). 
It is a long, but simple calculation to show that all these expansions agrees. 
Other exact results appear in 
Refs. \cite{s-92,sla-93,hz-93,f-92,slp-95,fz-96,wy-95,kh-07}
also for dynamical correlation functions.

\section{Conclusions}
\label{secconcl}

In this manuscript we proposed a variation of the replica method that allows 
the calculation of the large distance asymptotic of off-diagonal 
correlation functions in anyonic models. 
The main result we found is that the replicated correlation 
$Z_{2np}(t)$ is a function with $2p$ branches corresponding to different 
anyon statistics $\k=q/p$ with $q=0,\dots 2p-1$.
After having understood this, it is almost straightforward but tedious, 
to obtain all the harmonic terms of the large-distance expansion from 
known results for bosons. We applied the method to a gas of impenetrable
anyons (Tonks-Girardeau) and the most general anyonic Calogero-Sutherland.
For the former a number of analytic and numerical checks confirmed the
validity of the replica approach. For the latter instead, for special 
values of the interaction parameters (i.e. integer or rational), a 
number of exact results have been already found thanks to the random matrix
theory and its connection with Jack 
polynomials \cite{sla-93,hz-93,f-92,h-94,slp-95,fz-96}.
In some cases the correspondence of the two approaches is obvious,
while in some others (for rational $\l=p/q$ the correlation functions are
given by sum over fractional excitations involving $p+q$ quantum numbers,
giving an irregular function of $\l$ with no clear limit to irrational values) 
it implies particular relations for Jack polynomials. 
Showing in full generality this correspondence remains an open problem.

At the same time, another open problem is the correspondence of our approach 
to the one by Patu, Korepin and Averin \cite{pka-08} based on Fredholm 
determinants. We stress that (as far as we are aware) the
explicit correspondence of the two approaches has not
yet been proved even for bosons. 
The results coming for these two sets of works give an almost complete 
physical picture of the behavior of the off-diagonal correlator 
of the gas of impenetrable anyons. 
Beyond the impenetrable limit, only the Bethe solution of the Lieb-Liniger
model is known \cite{k-99,bg-06}, with still no attempt to tackle the hard
problem of the correlation functions exactly. 
Only the power-law structure (and the corresponding
singularities) are known from bosonization \cite{cm-07,pka-07}.
A powerful approach would be to generalize the quantum
inverse scattering methods for bosons \cite{Kbook} to anyons and then to mix 
integrability and numerics to get the full correlation functions from the 
form-factors (on the line of Ref. \cite{cc-06} for bosons).

\section*{Note Added}
After the completion of this work, a manuscript appeared where some large
distance asymptotic of the off-diagonal correlations for the Tonks-Girardeau 
gas have been found at finite temperature \cite{pka-new}. 
The results obtained there are different and complementary to ours.
Notice that the integral system of partial differential equations they derive
is independent on the anyonic parameter $\k$ that enters only through the
initial condition, on the line of our previous \cite{sc-08} and current 
results.

\section*{Acknowledgments}
We thank G. Marmorini, S. Ouvry, V. Pasquier, E. Tonni, and P. Wiegmann 
for useful discussions. 
This work has been done in part when the authors were guest of the Galileo
Galilei Institute in Florence, whose hospitality is kindly acknowledged.  
Financial support of the ESF (INSTANS activity) is also acknowledged.

\appendix

\section{The analytic continuation}

We need to analytic continue terms of the form 
\be
h_m^l=\prod_{a=1}^l \frac{\Gamma(a/\l)}{\Gamma((m-l+a)/\l)}\,.
\ee
Using the integral representation of the log of the Gamma function
\be
\log \Gamma(z) =\int_0^\infty\frac{dt}{t}\left( 
\frac{e^{-z t} - e^{-t}}{1 - e^{-t}} + (z - 1) e^{-t}\right)\,,
\ee
summing the resulting geometrical series and changing integration variable, 
we have 
\be\fl
\log h_m^l=  
\int_0^\infty\frac{dt}t e^{-t}\left[
\frac{1- e^{-tl}-e^{-t(m-l)}+e^{-m t}}{(1-e^{-t})(1-e^{-\l t})}-
\frac{(m-l)l}{\l} e^{-(\l-1)t}\right]\,.
\label{hm}
\ee
For the first harmonic and any $\k$ and $\l$, 
the single terms in the integral all diverge, only their sum is finite.
The integral is an analytic function of $l$ and $m$ and can then be used 
to evaluate $h_m^l$ for any complex $m,l$ for the first harmonic.
The other are obtained from  the recurrence relation in the text. 

For the coefficients $H_m^l(\l)$ in the text we also need 
the analytic continuation of $f_m=\Gamma((m+c)/\l)/\Gamma(c/\l)$ 
that are very similar
\bea
\log f_m&=& -\int_0^\infty \frac{dt}t e^{-t}
\left[
\frac{(1-e^{-m t})^2}{(1-e^{-t})(1-e^{-\l t})}-\frac{m}\l e^{-(\l-1)t} 
\right]\,,
\eea
that is vanishing for $\l=1$ and $m\to1$ as it should. 
Plugging everything together
\be
\log H_m^l(\l)=2\log h_m^l+\log f_m\,.
\ee


\section*{References}

\begin{thebibliography}{99}

\bibitem{anyons} 
J. Leinaas, J. Myrheim, Nuovo Cimento B {\bf 37}, 1(1977);
 F. Wilczek, Phys. Rev. Lett. {\bf 48}, 1144 (1982);
 F. Wilczek, Fractional Statistics and Anyon Superconductivity, 
(World Scientific, Singapore 1990).



\bibitem{1dexp}
H. Moritz, T. Stoferle, M. K\"ohl and T. Esslinger, 
Phys. Rev. Lett. {\bf 91}, 250402 (2003);
B. Paredes, A. Widera, V. Murg, O. Mandel, S. Folling, I. Cirac, 
G. V. Shlyapnikov, T. W. Hansch and I. Bloch, Nature 429, 277 (2004);
T. Kinoshita, T. Wenger and D. S. Weiss, Science {\bf 305}, 1125 (2004);
T. Kinoshita, T. Wenger and D. S. Weiss, Phys. Rev. Lett. {\bf 95}, 190406 (2005);
T. Kinoshita, T. Wenger and D. S. Weiss, Nature {\bf 440}, 900 (2006);
A. H. van Amerongen, J. J. P. van Es, P. Wicke, K. V. Kheruntsyan, 
and N. J. van Druten, Phys. Rev. Lett. {\bf 100}, 090402 (2008) [0709.1899].


\bibitem{par}
L. Jiang, G. K. Brennen, A. V. Gorshkov, K. Hammerer, M. Hafezi, E. Demler,
M. D. Lukin, and P. Zoller, Nature Phys. {\bf 4}, 482 (2008) [0711.1365];
B. Paredes, P. Fedichev, J. I. Cirac, and P. Zoller, 
Phys. Rev. Lett. {\bf 87}, 10402 (2001) [cond-mat/0103251];
M. Aguado, G. K. Brennen, F. Verstraete, and J. I. Cirac, 0802.3163.

\bibitem{k-99}
A. Kundu, Phys. Rev. Lett. {\bf 83}, 1275 (1999) [hep-th/9811247].

\bibitem{bg-06}
M. T. Batchelor, X. W. Guan, N. Oelkers, 
Phys. Rev. Lett. {\bf 96}, 210402 (2006) [cond-mat/0603643];
M. T. Batchelor, X. W. Guan, J.S.-He, 
J. Stat. Mech. (2007) P03007 [cond-mat/0611450];
M. T. Batchelor, X. W. Guan, 
Phys. Rev. B {\bf  74}, 195121 (2006) [cond-mat/0606353].

\bibitem{oae-99}
L. Amico, A. Osterloh, and U. Eckern, Phys. Rev. B {\bf 58} R1703 (1998)
[cond-mat/9803074];
A. Osterloh, L. Amico, and U. Eckern, J. Phys. A {\bf 33}, L87 (200)
[cond-mat/9812317];
J. Phys. A {\bf 33}, L487 (2000) [cond-mat/0007081];
Nucl. Phys. B {\bf 588}, 531 (2000) [cond-mat/0003099]. 

\bibitem{g-06} 
M. D. Girardeau, Phys. Rev. Lett. {\bf 97}, 210401 (2006) [cond-mat/0604357].

\bibitem{ssc-07}
R. Santachiara, F. Stauffer, and D. Cabra, 
J. Stat. Mech. (2007) L05003 [cond-mat/0610402].

\bibitem{zw-92}
J. Zhu and Z. D. Wang, Phys. Rev. A {\bf 53}, 600 (1992).

\bibitem{bg-06b}
M. T. Batchelor and X. W. Guan, 
Laser Phys. Lett. {\bf 4}, 77 (2007) [cond-mat/0608624].

\bibitem{cm-07}
P. Calabrese and M. Mintchev, 
Phys. Rev. B {\bf 75}, 233104 (2007) [cond-mat/0703117].

\bibitem{pka-07}
O. I. Patu, V. E. Korepin and D. V. Averin, 
J. Phys. A {\bf 40}, 14963 (2007) [0707.4520].

\bibitem{an-07}
D. V. Averin and J. A. Nesteroff, 
Phys. Rev. Lett. {\bf 99}, 096801 (2007) [0704.0439];

\bibitem{lm-99}
A. Liguori, M. Mintchev, and L. Pilo, 
Nucl. Phys. B {\bf 569}, 577 (2000) [hep-th/9906205];
A. Liguori and M. Mintchev, 
Commun. Math. Phys. {\bf 169}, 635 (1995) [hep-th/9403039].

\bibitem{it-99}
N. Ilieva and W. Thirring, 
Eur. Phys. J. C {\bf 6}, 705 (1999) [hep-th/9808103];
Theor. Mat. Phys. {\bf 121}, 1294 (1999) [math-ph/9906020].

\bibitem{kl-05}
E.-A. Kim, M. J. Lawler, S. Vishveshwara, E. Fradkin, 
Phys. Rev. Lett. {\bf 95}, 176402 (2005) [cond-mat/0507428];
Phys. Rev. B {\bf 74}, 155324 (2006) [cond-mat/0604325].

\bibitem{fibo}
A. Feiguin, S. Trebst, A. W. W. Ludwig, M. Troyer, A. Kitaev, Z. Wang, and 
M. H. Freedman, Phys. Rev. Lett. {\bf 98}, 160409 (2007) [cond-mat/0612341];
S. Trebst, E. Ardonne, A. Feiguin, D. A. Huse, A. W. W. Ludwig, 
M. Troyer, Phys. Rev. Lett. {\bf 101}, 050401 (2008) [0801.4602];
L. Fidkowski, G. Refael, N. Bonesteel, and J. Moore, 0807.1123.

\bibitem{g-07}
M. Greiter, arXiv:0707.1011.

\bibitem{zw-07}
R.-G. Zhu and A.-M. Wang, arXiv:0712.1264.

\bibitem{o-07}
S. Ouvry, arXiv:0712.2174.

\bibitem{sc-08}
R. Santachiara and P. Calabrese, J. Stat. Mech. (2008) P06005 [0802.1913].

\bibitem{pka-08}
O. I. Patu, V. E. Korepin and D. V. Averin, 
J. Phys. A {\bf 41}, 255205 (2008) [0801.4397];
J. Phys. A {\bf 41},  145006 (2008) [0803.0750].

\bibitem{hzc-08}
Y. Hao, Y. Zhang, and S. Chen, 
Phys. Rev. A {\bf 78}, 023631 (2008) [0805.1988].

\bibitem{dc-08}
A. del Campo,   Phys. Rev. A {\bf 78}, 045602 (2008) [0805.3786].

\bibitem{bgk-08}
M. Batchelor, X.-W. Guan, and A. Kundu, J. Phys. A {\bf 41} (2008) 
352002 [0805.1770].

\bibitem{bcm-08}
B. Bellazzini, P. Calabrese, and M. Mintchev, 0808.2719.



\bibitem{gk-01} D.~M.~Gangardt and A.~Kamenev, 
Nucl.~Phys.~B {\bf 610}, 578 (2001) [cond-mat/0102405]; 
S. M. Nishigaki, D. M. Gangardt, and A. Kamenev, 
J. Phys. A {\bf 36}, 3137 (2003) [cond-mat/0207301].

\bibitem{g-04}
D. M. Gangardt, J.Phys. A {\bf 37}, 9335 (2004) [cond-mat/0404104].

\bibitem{gs-06}
D. M. Gangardt, G. V. Shlyapnikov, 
New J. of Phys. {\bf 8}, 167 (2006) [cond-mat/0606319].

\bibitem{agls-06}
G.E. Astrakharchik, D.M. Gangardt, Yu.E Lozovik, and I.A. Sorokin,
Phys. Rev. E {\bf 74}, 021105 (2006) [cond-mat/0512470].

\bibitem{sg}
M. Mezard, G. Parisi, M. A. Virasoro, Spin glass theory and beyond. Singapore:
World Scientific (1987).





\bibitem{LL}
E. H. Lieb and W. Liniger, Phys. Rev. {\bf 130}, 1605 (1963);
E. H. Lieb, Phys. Rev. {\bf 130}, 1616 (1963).

\bibitem{k-91} J.~Kurchan, J. Phys. A {\bf 24}, 4969 (1991).

\bibitem{fw-04} 
P.J.~Forrester and N.S.~Witte, Nagoya Math. J. {\bf 174}, 29 (2004).

\bibitem{h-81}
F. D. M. Haldane, Phys. Rev. Lett. {\bf 47}, 1840 (1981);
J. Phys. C {\bf 15}, 2585 (1981).

\bibitem{metha}
M. L. Metha, Random Matrices, (Academic Press, New York, 1991).

\bibitem{df}
V. S. Dotsenko and V. A. Fateev, Nucl. Phys. B {\bf 251} (1985), 691.

\bibitem{fz-96} 
P. J.~Forrester and J. A.~Zuk, Nucl.~Phys.~B {\bf 473}, 616 (1996).

\bibitem{fw-02}
P. J. Forrester and N. S. Witte, 
Commun. Pure Appl. Math. {\bf 55}, 679 (2002) [math-ph/0204008].

\bibitem{ffgw-03}
P.J. Forrester, N.E. Frankel, T.M. Garoni, and N.S. Witte, 
Commun. Math. Phys. {\bf 238}, 257 (2003) [math-ph/0207005].

\bibitem{jmms-80}
M. Jimbo, T. Miwa, Y. Mori, and M. Sato, Physica D {\bf 1}, 80 (1980).

\bibitem{vt-79}
H. G. Vaidya and C. A. Tracy, 
Phys. Rev. Lett. {\bf 42}, 3 (1979) [Phys. Rev. Lett. 43, E1540 (1979)];
J. Math. Phys. {\bf 20}, 2291 (1979).

\bibitem{ctw-80}
D. B. Cramer, H. B. Thacker, and D. Wilkinson,
Phys. Rev. D {\bf 23}, 3081 (1981).



\bibitem{s-71} B. Sutherland, J. Math. Phys. \textbf{12}, 246
  (1971); \textbf{12}, 251 (1971); Phys. Rev. A \textbf{4}, 2019 (1971);
  \textbf{5}, 1372 (1972).

\bibitem{h-94} Z. N. C.~Ha, Phys.~Rev.~Lett. {\bf 73}, 1574 (1994);
Nucl. Phys. B {\bf 435}, 604 (1995).

\bibitem{s-92}
B. Sutherland, Phys. Rev. B {\bf 45}, 907 (1992).

\bibitem{sla-93} B. D.~Simons, P. A.~Lee, and B. L.~Altshuler,
  Phys.~Rev.~Lett. {\bf 70}, 4122 (1993); 
   Nucl.~Phys. B {\bf 409}, 487 (1993).

\bibitem{hz-93} F. D. M.~Haldane and M. R.~Zirnbauer,
  Phys.~Rev.~Lett. {\bf 71}, 4055 (1993).

\bibitem{f-92} P. J.~Forrester, 
Nucl.~Phys.~B {\bf 388}, 671 (1992); 
Phys.~Lett.~A {\bf 179}, 127 (1993).

\bibitem{slp-95}
D.~Serban, F.~Lesage, V.~Pasquier, 
Nucl. Phys. B {\bf 435}, 585 (1995) [hep-th/9405008];
Nucl.~Phys.~B {\bf 466}, 499 (1996) [hep-th/9508115].



\bibitem{wy-95}
Y.-S. Wu and Y. Yu, Phys. Rev. Lett. {\bf 75}, 890 (1995).

\bibitem{kh-07}
H. Katsura and Y. Hatsuda, J. Phys. A {\bf 40}, 13931 (2007).









































\bibitem{Kbook}
V. E. Korepin, N. M. Bogoliubov and A. G. Izergin, 
Quantum Inverse Scattering Method and Correlation Functions 
(Cambridge Un. Press, Cambridge, 1993), and references therein.

\bibitem{cc-06}
J.-S. Caux and P. Calabrese, 
Phys. Rev. A {\bf 74}, 031605 (2006) [cond-mat/0603654];
 J.-S. Caux, P. Calabrese, and N. A. Slavnov, 
J. Stat. Mech. P01008 (2007) [cond-mat/0611321].

\bibitem{pka-new}
O. I. Patu, V. E. Korepin and D. V. Averin, 0811.2419.


\end{thebibliography}
\end{document}